\documentclass{appolb}
\usepackage{graphicx}

\preprint{}

\begin{document}

\title{How much is RHIC different from SPS? \\ Comparison of the  
$p_\perp$-spectra\thanks{Supported by the Polish State Committee for
Scientific Research, grant 2 P03B 09419.}}
\author{Wojciech Broniowski and Wojciech Florkowski 
\address{The H. Niewodnicza\'nski Institute of Nuclear Physics\\
PL-31342 Krak\'ow, Poland}
}
\maketitle

\begin{abstract}
We show, by means of a simple compilation of the available experimental
results, that the $p_\perp$-spectra obtained at RHIC and SPS are
strikingly similar up to $p_\perp \simeq 1.5-2$ GeV. In fact, the discrepancies between various
experimental groups working at the same experiment are of the same
size as the differences between RHIC and SPS. Our observation is
complementary to the well known fact of the equality of the measured
$R_{\rm side}$ and $R_{\rm out}$ HBT radii at RHIC and SPS. In
essence, it points out that the transverse size of the firecylinder
and the strength of the transverse flow are not significantly changed
between SPS and RHIC. This suggests that a saturation mechanism is effective
already at SPS. We also point out that the dominance of protons over 
$\pi^+$ at large $p_\perp$ can be seen not only in the RHIC data, but 
already in the SPS data.
\end{abstract}

 
\thispagestyle{empty}

\section{Introduction}

The pertinent question in the field of relativistic heavy-ion
collisions is whether the physics at RHIC is {\em qualitatively
different} from the physics at SPS. In our opinion, the available
experimental results hint that this is not the case, at least for soft
processes.

In this paper we compile the transverse-momentum spectra of hadrons
measured by various groups at SPS and RHIC. Surprisingly, 
to our knowledge such a study has not been presented before.
We use the data of NA44
\cite{na44}, NA49 \cite{na49}, PHENIX \cite{phenix}, and STAR
\cite{stara,starb,starc}, and show that there exist discrepancies
between the NA44 and the NA49 data, as well as between the PHENIX and
the STAR data. In fact, these discrepancies are of the same size as
the differences between SPS and RHIC. More precisely, within the experimental
uncertainties, which are quite large, one finds that the slopes of the
$p_\perp$-spectra at RHIC are compatible with those at SPS. This
observation is complementary to the well known fact of the very weak
beam-energy dependence of the transverse HBT radii. Similarity of both
the transverse size and the \hbox{$p_\perp$-spectra} of hadrons
indicates that the amount of the transverse flow cannot be
significantly different at the two considered collision energies.

The transverse HBT radii $R_{side}$ and $R_{out}$ measured \cite{phenixhbt,starhbt,agshbt}
in the first run of RHIC are very close to those measured in
heavy-ion collisions at smaller beam energies. Only the longitudinal
radius $R_{long}$ exhibits a monotonic growth with $\sqrt{s_{NN}}$
(for a compilation of the data at different energies see, e.g.,
Ref. \cite{phenixhbt}). The weak energy dependence of the transverse
radii is surprising, since the RHIC beam energy, $\sqrt{s_{NN}}=130$
GeV, is almost one order of magnitude larger than the SPS energy,
$\sqrt{s_{NN}}$ =17 GeV, and one would naively expect that much larger
hadronic systems were produced at RHIC. One would also expect a longer
lifetime of the hadronic fireball formed at RHIC, which should be
reflected in a longer emission times of pions. This effect is
quantified by the measurement of the ratio $R_{out}/R_{side}$, which
is expected to be much larger than unity for long emission duration
\cite{dumitru}.  The experimental measurements indicate, however,
that $R_{out}/R_{side}$ is compatible with unity in the whole range of
the studied transverse-momentum range ($ 0.2 < k_T < 1.0 $ GeV). This
fact is another puzzle delivered by the analysis of the RHIC data.

On the other hand, the first measurements at RHIC showed that the
pseudorapidity densities of charged particles are higher than those
observed at SPS. Can this effect be reconciled with practically
unchanged transverse radii?  For the most central collisions, PHOBOS
communicated the value $\langle N_{ch} \rangle = 555 \pm 12 \pm 35$
\cite{phobosmul}, BRAHMS the value $\langle N_{ch} \rangle = 549 \pm 1 \pm 35$, 
whereas PHENIX obtained $\langle N_{ch} \rangle = 622 \pm 1 \pm 41$ \cite{phenixmul}. 
Normalizing per participant pair yields $\langle N_{ch} \rangle /(0.5 \langle N_{part} \rangle) 
= 3.2 $ for both PHOBOS and BRAHMS, and $\langle N_{ch} \rangle /(0.5 \langle N_{part} \rangle) 
= 3.6 $ for PHENIX. These numbers may be compared to
the NA49 result, 1.9 \cite{na49mul,phobosmul}, and the WA98 result, 2.6
\cite{wa98}. We can see that the multiplicity increases by about 50\%
when we move from SPS to RHIC. A simple geometric scaling suggests
that the transverse radius increases, correspondingly, as a square root, 
{\em i.e.}, by about 20\%.  Thus,
the observed increase of the multiplicity translates to moderately small
increase of the transverse radii. Clearly, the difference between the
NA49 multiplicity and the WA98 multiplicity, as well as the
errors of each particular experiment lead to the uncertainty
in the determination of the geometric parameters. 

The weak dependence of $R_{side}$ from SPS to RHIC means that the
transverse size of the firecylinder changes very little. If
this is the fact, than the amount of the transverse hydrodynamic flow
should also be similar, since it is difficult to imagine that a much
stronger flow would lead to the same transverse size at freeze-out. To
the contrary, the STAR results for the $p_\perp$-spectra have been
interpreted as an indication of a much stronger flow at RHIC. Below we
redo the simple analysis of the hadronic spectra and show that the
errors in the flow parameter are very large, such that one cannot
definitely conclude that the flow is stronger at RHIC.  Moreover, the
PHENIX data suggest a much lower flow than STAR, such that it becomes
compatible with the flow at SPS, as obtained from NA44 and NA49.

\section{Compilation of the $p_\perp$-spectra measured at 
SPS and RHIC}

\begin{figure}[p]
\begin{centering}
\includegraphics[width=12.6cm]{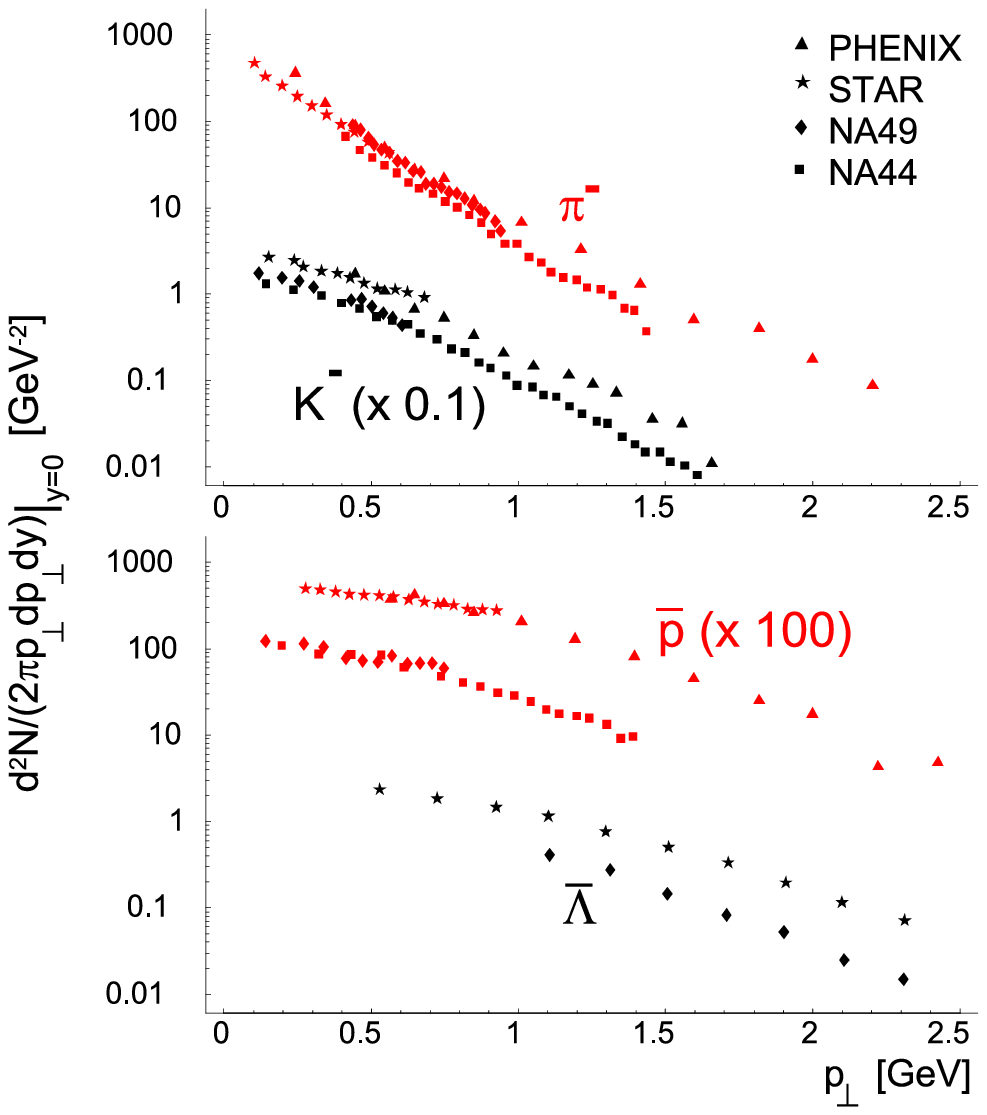}
\end{centering}
\label{fig1}
\caption{Comparison of the mid-rapidity $p_\perp$-spectra of $\pi^-$,
$K^-$, $\overline{p}$, and $\overline{\Lambda}$ for the most central
collisions of $Pb+Pb$ at $\sqrt{s_{NN}}=17$ GeV (NA44, NA49) and $Au +
Au$ at $\sqrt{s_{NN}}=130$ GeV (PHENIX, STAR). The STAR data for
$\pi^-$ and $K^-$, and the NA49 data are preliminary.}
\end{figure}

\begin{figure}[p]
\begin{centering}
\includegraphics[width=12.6cm]{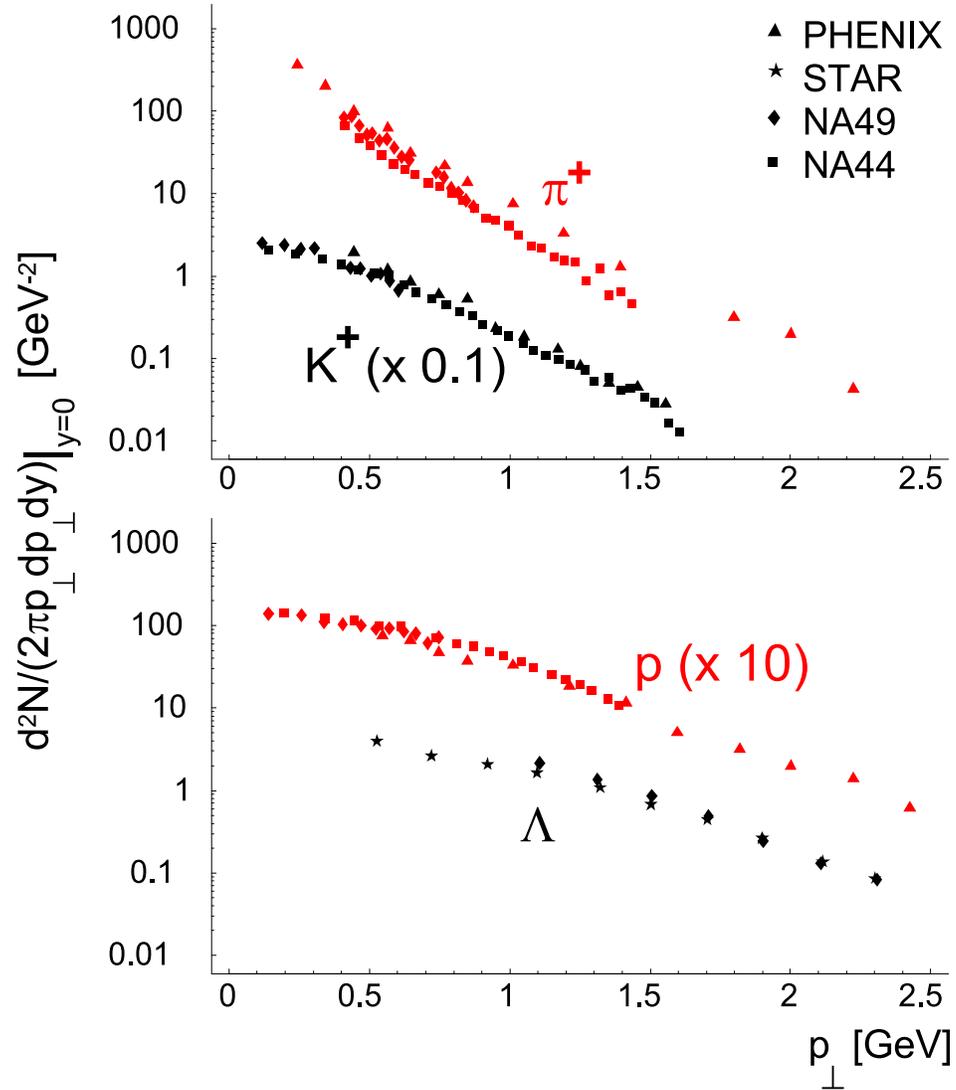}
\end{centering}
\label{fig2}
\caption{Same as Fig. 1 for $\pi^+$, $K^+$, $p$, and $\Lambda$.}
\end{figure}

\begin{figure}[p]
\begin{centering}
\includegraphics[width=12.6cm]{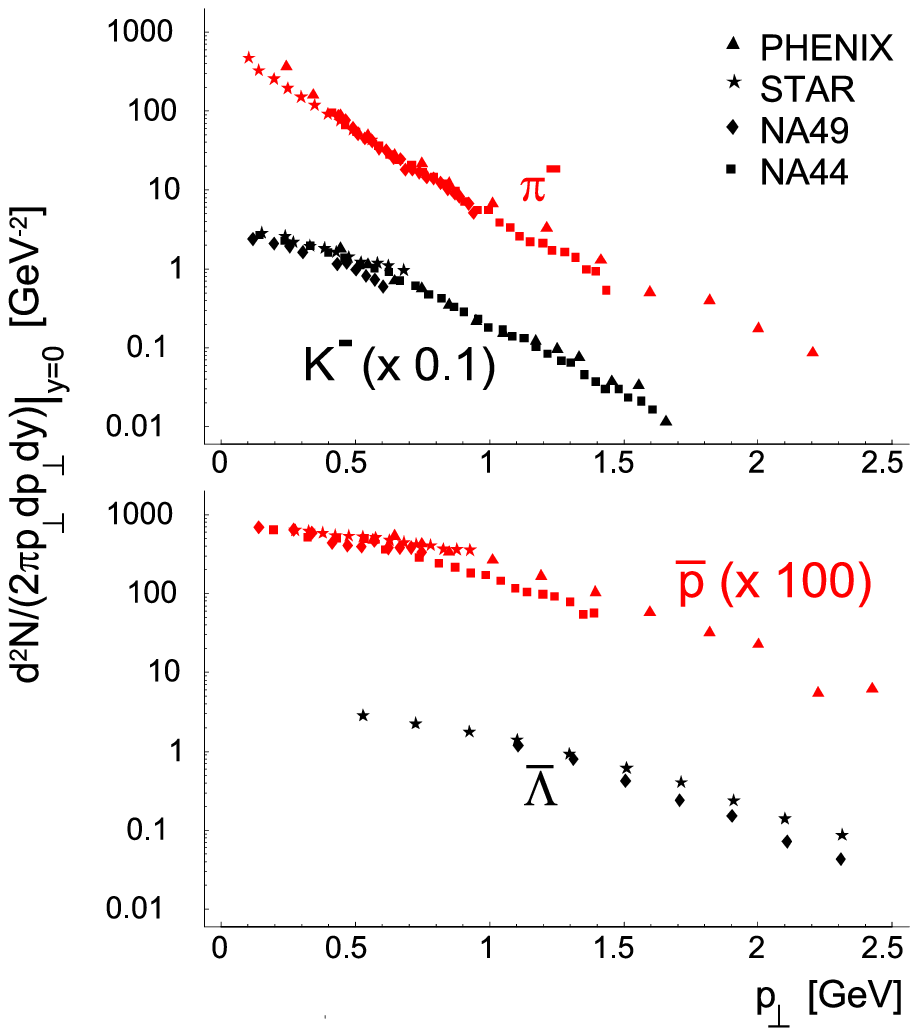}
\end{centering}
\label{fig3}
\caption{Same as Fig. 1 for the rescaled spectra.}
\end{figure}

\begin{figure}[p]
\begin{centering}
\includegraphics[width=12.6cm]{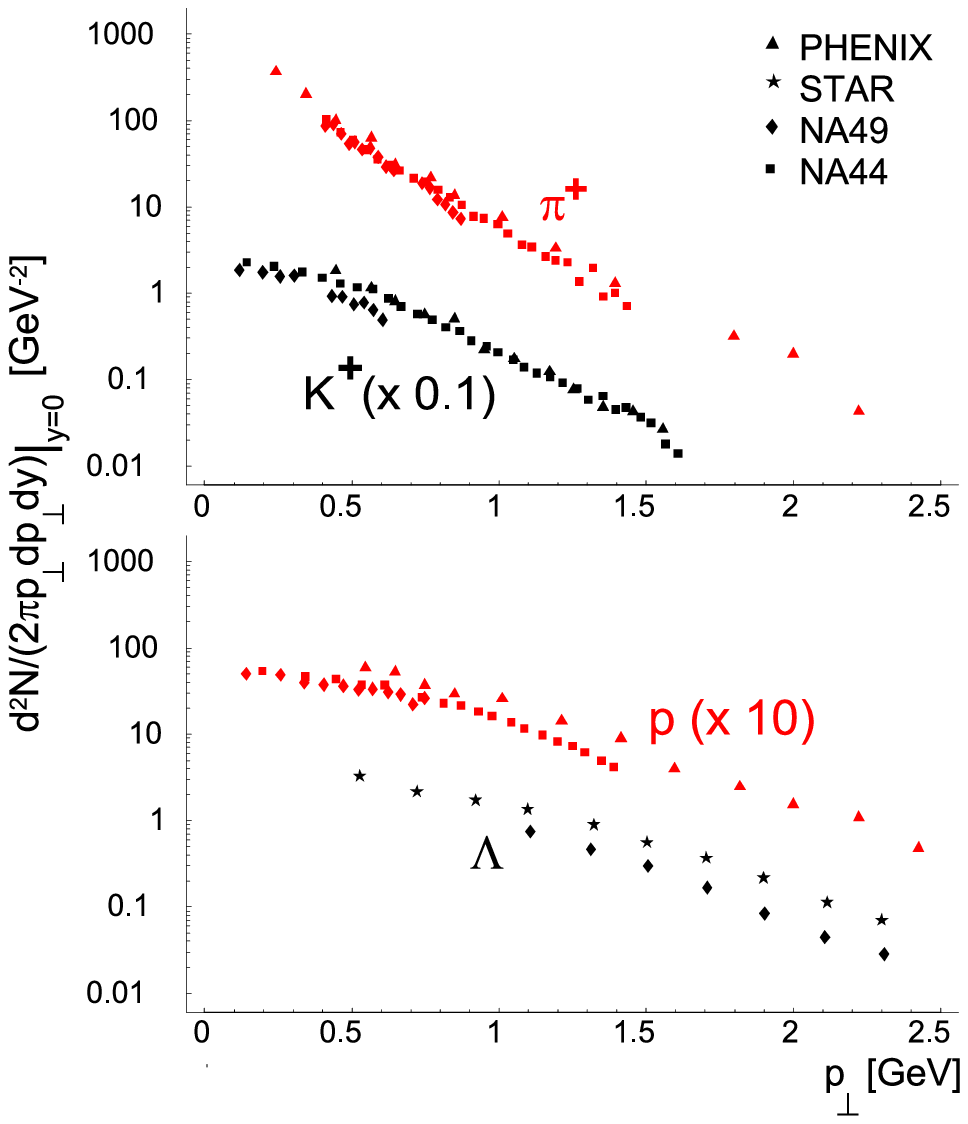}
\end{centering}
\label{fig4}
\caption{Same as Fig. 2 for the rescaled spectra}
\end{figure}

We begin by just displaying the experimental results from various
groups in a single plot.  In Fig.~1 we plot the $p_\perp$-spectra of
$\pi^-$, $K^-$, $\overline{p}$, and $\overline{\Lambda}$. Similarly,
in Fig.~2 we show the $p_\perp$-spectra of $\pi^+$, $K^+$, $p$, and
$\Lambda$.  The collected data come from the measurements done at
midrapidity for the most central events \cite{na44,na49,phenix,stara,starb,starc}.  
A striking feature of Fig.~1
is a very impressive agreement between different experiments for the
pions. In the range 0.4 $< p_\perp <$ 1.0 GeV, the data from NA49
coincide with the data from PHENIX and STAR, whereas the data from
NA44 show the same slope with a slightly smaller normalization due to
a different centrality choice (see the discussion below).  A very
similar $p_\perp$-dependence is also seen in the spectra of kaons,
$\overline{p}$, and $\overline{\Lambda}$ measured by different
experiments. In this case, however, a different normalization of the
spectra between SPS and RHIC data is clearly seen.
In Fig. 2 one can clearly see similarities in the shapes of the pion as well as 
the kaon spectra measured by NA44 and PHENIX.

If the plots for mesons and baryons were overlaid
({\em cf.} \cite{wbwf,hirsch}), one could see that the PHENIX
data have the property that the $p_\perp$-spectra of $\pi^+$ and $p$ cross around 
$p_\perp=2$ GeV, such that there are more protons than $\pi^+$ at large 
momenta. 
We wish to stress that the same phenomenon can be seen already at SPS, 
where the protons dominate $\pi^+$ 
at $p_\perp > 1$ GeV. The lower value of the crossing point reflects a much higher 
proton density at SPS compared to RHIC. 

At RHIC, the ``anomalous'' \cite{phenix,vitev} feature of the spectra is the dominance of 
$\overline{p}$ over $\pi^-$ at $p_\perp > 2$ GeV. 
One could speculate that a similar behavior might occur already at SPS, if the measurements 
had beed carried out to sufficiently high momenta. Indeed, when the NA44 data are extrapolated
with simple exponential functions, then one finds that a crossing occurs around $p_\perp=2.5$ GeV.
Since the exponential fits may not work over the large range in $p_\perp$, this phenomenon
remains a speculation till verified experimentally.  

In order to see the similarities between the data even more vividly, we scale the
normalization of the spectra in the following way:
\begin{enumerate}
\item In the NA49 data we undo the corrections for the feeding of
protons and antiprotons from weak decays. According to Ref. \cite{na49},
this correction is about 30\%, thus we divide the NA49 data for $p$
and $\overline{p}$ by the factor 0.7.

\item The most central data from NA44 correspond to the average impact
parameter $\langle b \rangle$ = 5 fm, whereas the most central NA49
data correspond to $\langle b \rangle$ = 2 fm \cite{na44}. This
difference explains a smaller normalization of the NA44 data. We
correct for the centrality choice of NA44, multiplying all NA44
spectra by an educated-guess factor of 1.5.

\item Inspired by the success of the thermal approach, we scale the
spectra by the factor $\exp[-(B \mu_B + S \mu_S + I_3 \mu_I)/T]$,
where $T$ is the freeze-out temperature, $B$, $S$, and $I_3$ are the
baryon number, strangeness, and the third component of the isospin of
the particle, respectively, and $\mu_B$, $\mu_S$, and $\mu_I$ are the
corresponding chemical potentials. At SPS ($Pb+Pb$, $\sqrt{s_{NN}}=17$
GeV) \cite{michalec}
\begin{equation}
T=164 {\rm ~MeV}, \;\; \mu_B=229{\rm ~MeV}, \;\;  \mu_S=54{\rm ~MeV}, 
\;\; \mu_I=-7{\rm ~MeV},
\label{spspar}
\end{equation}
while at RHIC ($Au+Au$, $\sqrt{s_{NN}}=130$ GeV) \cite{fbm}
\begin{equation}
T=165 {\rm ~MeV}, \;\; \mu_B=41{\rm ~MeV}, \;\;  \mu_S=9{\rm ~MeV}, 
\;\; \mu_I=-1{\rm ~MeV}.
\label{rhicpar}
\end{equation}
Since in
the thermal model the (original) spectra are proportional to the
factor $\exp[(B \mu_B + S \mu_S + I_3 \mu_I)/T]$ (for the Boltzmann
statistics, which works very well), our rescaling approximately
removes the effects of different chemical potentials at RHIC and
SPS. Due to resonance decays, even within the thermal model the
scaling is not exact, but approximate.
\footnote{For instance, a larger baryon chemical potential leads to a larger 
pion yield from the process $\Delta \to \pi N$, hence the secondary pions are
sensitive to the baryon chemical potential.}
\end{enumerate}

\begin{figure}[p]
\includegraphics[width=12.6cm]{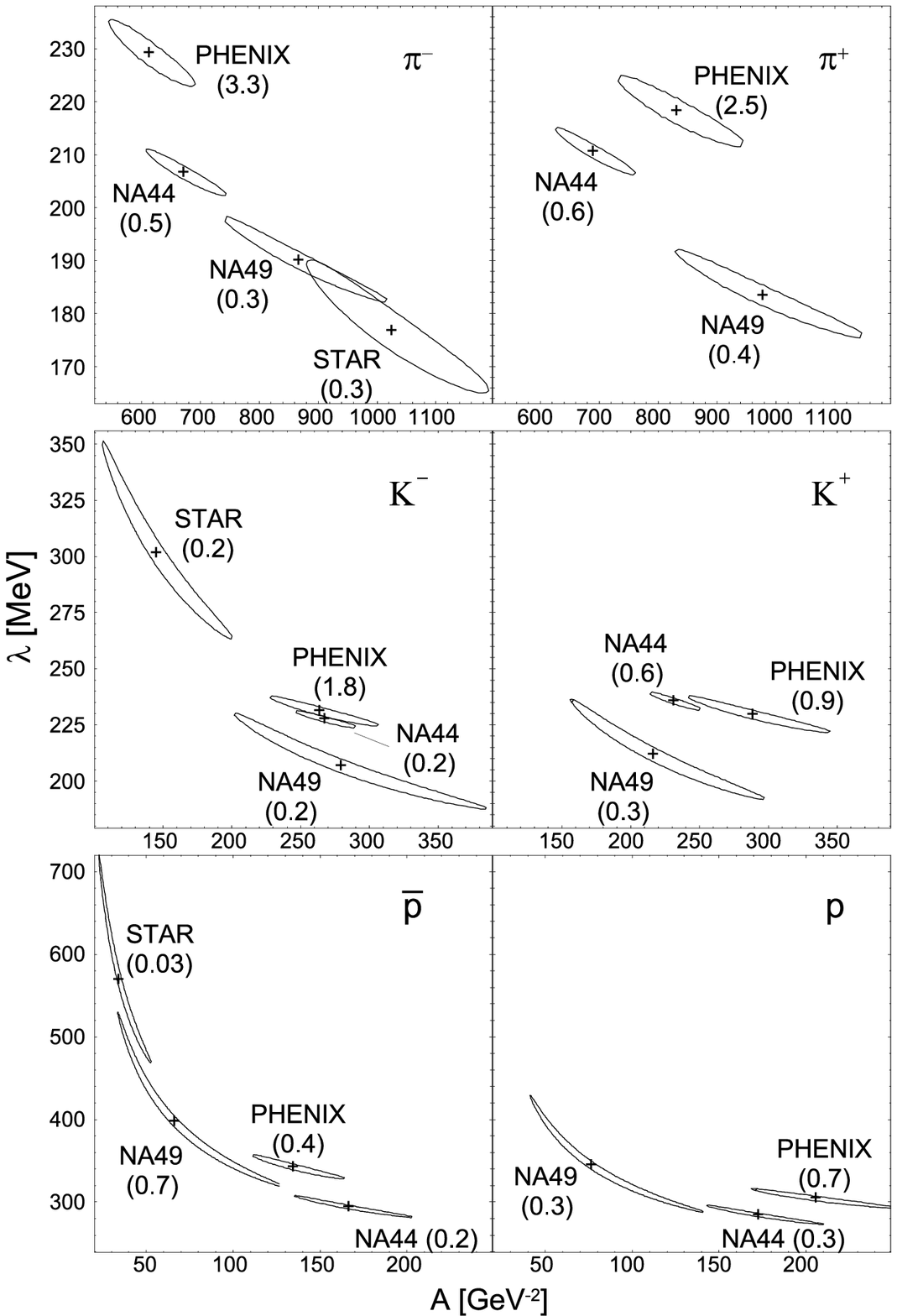}
\label{chi2}
\caption{The contours of $\Delta \chi^2=1$ for various fits of the
rescaled mid-rapidity \hbox{$p_\perp$-spectra} to the exponential form $A
\exp (-m_\perp/\lambda)$. Optimum values are denoted by crosses.  The
numbers in parentheses are the values of the $\chi^2$ per degree of
freedom at the optimum.  The data for NA49 and the STAR data for
$\pi^-$ and $K^-$ are preliminary. }
\end{figure}

The described rescaling is not necessary for our analysis, and has no
impact on the conclusions. However, it is useful, since it brings the
spectra closer and makes the eye-ball comparison easier.  The results
are shown in Figs. 3 and 4.  For the pions, the four measurements
agree very well in the overlap region. For the kaons, the NA49 data
have a visibly larger slope, while the three other sets of data
overlap. The measurements for protons and $\overline{p}$ agree 
as well. The spectra of $\Lambda$ and
$\overline{\Lambda}$ have smaller slopes in the STAR measurement
compared to NA49. One should bare in mind that the
statistical errors are typically of the order of a few percent, and
the systematic errors are around 10\%. Hence, within the experimental
errors, there are no significant differences between SPS and RHIC
which could indicate different physics. Moreover, the small difference
between NA44 and NA49, and between PHENIX and STAR, which one may
observe in Figs. 3 and 4 is of the same magnitude as the difference
between SPS and RHIC.

\section{Exponential fits}

In this Section we wish to quantify the eye-ball observations from Figs.~3 and
4.  The most popular method is to fit the spectra to the exponential
function, and compare the parameters. In fact, this is the most economic and common
way of presenting the data, and the differences in the slope
parameters of particles of different masses are interpreted as the
signature of the transverse flow \cite{bl,scheibl}.  Hence, we fit the
function\footnote{Other forms used in the literature, differing by the power
of $m_\perp$, lead to similar qualitative conclusions.}
\begin{equation}
\left . \frac{d^2N}{m_\perp d m_\perp dy} \right |_{y=0} = 
A \exp(-m_\perp / \lambda)
\label{expo}
\end{equation}
to each {\em rescaled} spectrum, independently for different
experimental groups.\footnote{We note that the slope parameters, $\lambda$, fitted
in many papers, depend strongly on the choice of the range in
$p_\perp$ (see, e.g., the discussion in \cite{strange}). Here we use
the available data ranges with $p_\perp < 2$ GeV. Because of the lack of strict
thermodynamic interpretation, we refrain from calling $\lambda$ the temperature
of the spectrum.}  The method is as
follows: we constrain the fit to $p_\perp < 2$GeV (this has relevance
for the PHENIX data, and for $\Lambda$ and $\overline{\Lambda}$ measured by STAR). 
For simplicity, we assume a 15\% error on every
point. This is in the ballpark of the errors given by the
experimental groups. Then, the $\chi^2$ function is minimized with
respect to the $A$ and $\lambda$ parameters.

The results are shown in Fig. 5. The first feature to notice are large
correlations between $\lambda$ and $A$, which lead to
sizeable errors in these parameters. The optimum values for $\lambda$, denoted
in Fig.~5 by crosses, and the errors are listed in Table \ref{invsl}.  While the optimum
values agree with the estimates made in other papers
\cite{na44,na49,phenix,stara}, our errors are significantly larger for
the case of STAR and NA49.  The physically relevant parameter is
the inverse slope, $\lambda$, while the norm parameter, $A$, carries
the ambiguities described above (rescaling).  Hence, if the results of
the two data sets are to be consistent, then the ranges of $\lambda$
(and not necessarily $A$) should overlap. We note from Fig. 5
that this is basically the case for NA44 and NA49, except perhaps the
case of $\pi^+$. The STAR and PHENIX data are much less consistent,
especially for the case of $\pi^-$.  For kaons, protons, and
$\overline{p}$, the PHENIX data agree within errors with NA44 and
NA49.  Some caution is needed in the interpretation of Fig.~5,
since the fits are made over different ranges in $p_\perp$, and the
form used for the fit has no sound physical ground. Yet, there are
indications for discrepancies between STAR and PHENIX. Also, it is
clear that only the STAR data suggest larger values of $\lambda$ (albeit with
large errors), which would mean larger transverse flow. If PHENIX data are used,
no such conclusion comes out, since the PHENIX data are basically
consistent with SPS. Note also that the quality of the fit, indicated by the
value of $\chi^2$ at the optimum, is worst for the case of pions from PHENIX. This reflects
the inapplicability of the simple exponential parameterization over the large range in 
$p_\perp$.

\begin{table}
\begin{center}
\begin{tabular}{|l|llll|}
\hline
 & NA44 & NA49 & PHENIX & STAR \\
\hline
$\pi^+$ & $211 \pm 4$           & $183 \pm 8$   & $218 \pm 7$   &               \\
$\pi^-$ & $207 \pm 4$           & $190 \pm 8$   & $229 \pm 6$   & $176 \pm 12$  \\
$K^+$   & $235 \pm 4$           & $211 \pm 22$  & $230 \pm 8$   &               \\
$K^-$   & $227 \pm 4$           & $207 \pm 21$  & $231 \pm 7$   & $301 \pm 43$  \\
$p$     & $284 \pm 12$          & $344 \pm 68$  & $304 \pm 12$  &               \\
$\overline{p}$  & $294 \pm 14$  & $397 \pm 100$ & $342 \pm 15$  & $569 \pm 122$ \\
${\Lambda}$ &                   & $294 \pm 24 $ &               & $372 \pm 22$  \\
$\overline{\Lambda}$ &          & $298 \pm 27 $ &               & $398 \pm 25$  \\
\hline
\end{tabular}
\caption{Our values for the inverse-slope parameters, $\lambda$, in units of MeV, fitted to mid-rapidity 
data from the most-central collisions 
of $Pb+Pb$ at $\sqrt{s_{NN}}=17$ GeV (NA44 \cite{na44}, NA49 \cite{na49}) and $Au + Au$ at 
$\sqrt{s_{NN}}=130$ GeV (PHENIX \cite{phenix}, STAR \cite{stara,starb,starc}). The fit includes 
data points up to 2 GeV. 
The NA49 data, and the STAR data for $\pi^-$ and $K^-$ are preliminary.}
\label{invsl}
\end{center}
\end{table}

The final remark of this Section concerns the multiplicities of particles, 
\begin{equation}
\left . \frac{dN}{dy} \right |_{y=0} = 
\int_0^\infty \left . dp_\perp p_\perp \frac{d^2N}{p_\perp d p_\perp dy} \right |_{y=0}.
\label{ppp}
\end{equation}
The integrand in Eq. (\ref{ppp}) peaks at $p_\perp \simeq 0.25$ GeV for the pions, and 
0.6 GeV for the protons. Therefore much of the strength comes from the $p_\perp$ region
not covered by the data. The necessary extrapolation brings systematic uncertainties to
$dN/dy$ and may be a source of discrepancies between various quoted numbers.

\section{A model calculation}

The estimates of the transverse flow must be based on model
calculations which include this effect, as well as other potentially
important physical effects.  In this Section we apply the {\em
single-freeze-out model}, which has been used successfully to describe
the $p_\perp$-spectra at RHIC. The model has been described in detail
in of Refs. \cite{wbwf,hirsch,strange}, and here we only list its basic
features: i) simultaneous chemical and thermal freeze-out of the
hadronic matter, ii) inclusion of all hadronic resonances, and iii) a
simple parametrization of the freeze-out hypersurface, which is
defined by the condition $\tau=\sqrt{t^2-x^2-y^2-z^2}=$const. The
hydrodynamic flow on the freeze-out hypersurface is taken in the form
resembling the Hubble law, i.e., $u^\mu=x^\mu/\tau$. That way both the
longitudinal and transverse flows are built in.  The single-freeze-out
model has two thermodynamic parameters (temperature and the baryon
chemical potential) which are fixed by the global fit to the relative
particle yields ({\em cf.} Eqs. (\ref{spspar},\ref{rhicpar})).  The
extra two parameters ($\tau$ and the transverse radius of the
firecylinder $\rho_{\rm max}$) determine the overall normalization and
the shape of the spectra. In our present calculation the thermodynamic
parameters are the same as those used previously in 
Sect.~2.~\footnote{We note that $T$ and $\mu_B$ are the
only independent thermodynamic parameters, since $\mu_S$ and $\mu_I$
follow from the conservation of strangeness and electric charge.}
Then, the expansion parameters $\tau$ and $\rho_{\rm max}$ are fixed
separately to the data of each experimental group. They determine, in
each particular case, the maximal transverse flow, given by the model
formula
\begin{equation}
\beta^{\rm max}_\perp = { \rho_{\rm max} \over \sqrt{\tau^2 +
\rho_{\rm max}^2}}.
\label{bmax}
\end{equation}
The average value of the transverse flow velocity, $\langle
\beta_\perp \rangle$, is very close to $(2/3) \; \beta^{\rm max}_\perp$.

\begin{figure}[p]
\begin{centering}
\centerline{\includegraphics[width=10.7cm]{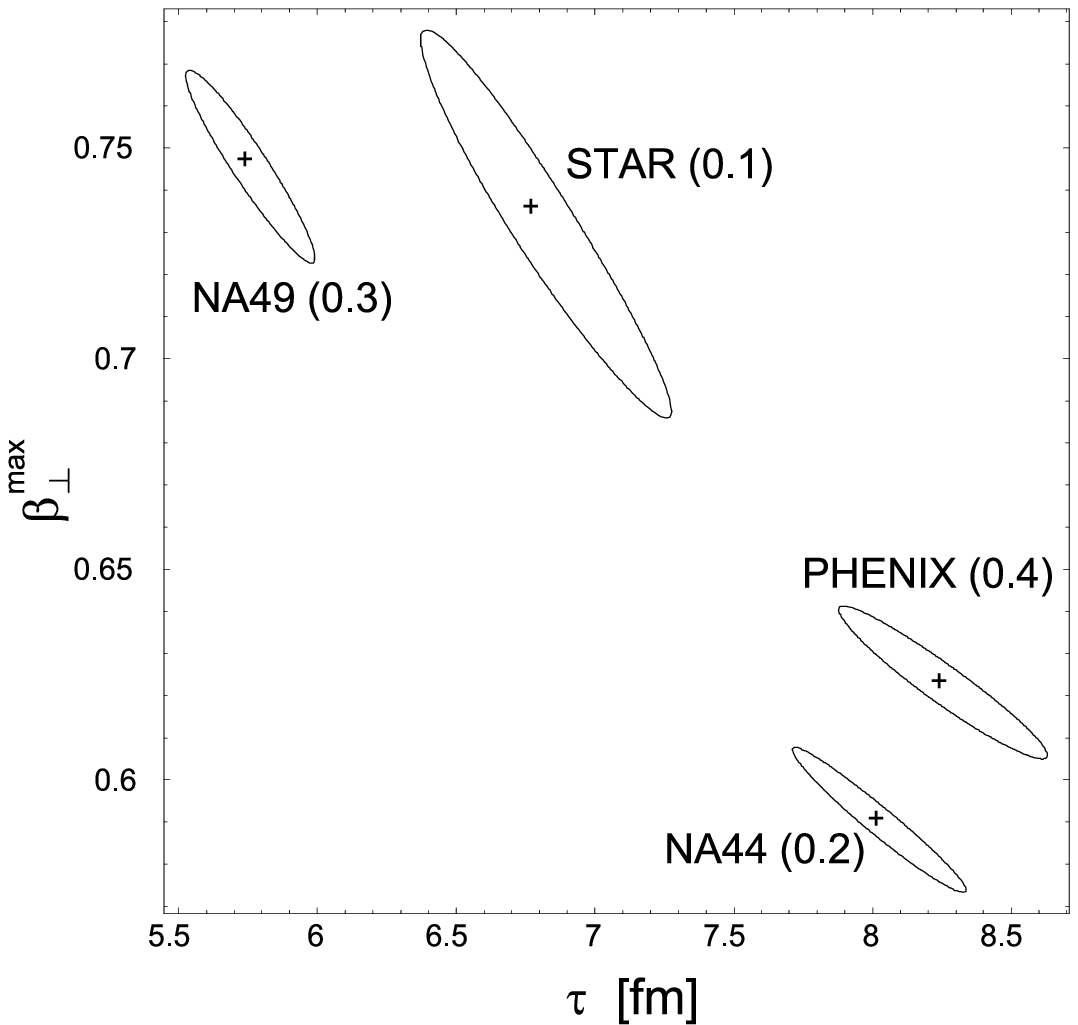}}
\end{centering}
\label{mod}
\caption{The contours of $\Delta \chi^2=1$ for fits of the combined
mid-rapidity $p_\perp$-spectra to the {\em single-freeze-out model} of
Refs. \cite{wbwf,strange}. The parameters $\tau$ and $\beta_\perp^{\rm
max}$ are defined in the text.  The optimum values of the parameters
are denoted by crosses.  The numbers in parentheses are the values of
the $\chi^2$ per degree of freedom at the optimum.  The NA49 data and
the STAR data for $\pi^-$ and $K^-$ are preliminary.}
\end{figure}

The results are presented in Fig.~6, where we show the optimum values
for the invariant time, $\tau$, and the flow parameter, $\beta^{\rm
max}$, together with their errors.  If the results were consistent,
then the fitted values would overlap. This is not the case.  The fit
to NA44 is far away from the fit to NA49, and the fit to STAR is far
away from the fit to PHENIX. Amusingly, PHENIX is close to NA44, and
STAR is close to NA49. The grand average of $\beta^{\rm max}$ from
combined PHENIX and STAR is close to the average from combined NA44
and NA49 experiments.  These results indicate that even within a model
capable of explaining the spectra one cannot conclude of a larger
transverse flow at RHIC compared to SPS.

\section{Conclusions}

Here are our main points:

\begin{enumerate}

\item The similarity of $p_\perp$-spectra at SPS and RHIC in the data
range suggests similar soft physics. We have argued that the combined
present data do not lead to the conclusion of much larger transverse
flow at RHIC. Only the STAR data for $\overline{p}$ and $K^-$ support
this view, although with large experimental uncertainty. The PHENIX
data are compatible with the same transverse flow as in SPS.

\item The property of the spectra that there are more protons
than $\pi^+$ at large $p_\perp$ can be seen not only in
the RHIC data, but also in the SPS data.

\item At the moment the experimental discrepancies between NA44 and
NA49, and between STAR and PHENIX are of the same magnitude as the
discrepancies between SPS and RHIC.  With better data more accurate
conclusion would be achieved. In particular, the RHIC measurements at
lower energies will be very useful in verifying theoretical
hypotheses.  The use of models failing to reproduce the RHIC data
should be, if possible, avoided in the modeling of detectors and in
the analysis of the data.  Also, the frequently made corrections for
weak decays are not very useful, since these can be accounted for
without difficulty in theoretical models. When comparing the corrected
data much care is needed as to how the feeding from weak decays has
been subtracted.

\item We recall that the freeze-out temperatures used in the framework
of thermal models have saturated (compare Eqs. (\ref{spspar}) and
(\ref{rhicpar})), which was not yet the case at AGS energies.  Note,
however, that the thermal approach works well for the AGS data
\cite{PBMAGS,cest} and for elementary collisions \cite{becattini}.

\item The transverse HBT radii and the slopes of the spectra are
similar at SPS and RHIC, which implies similar transverse flow, or, in
general, similar soft transverse physics. The particle yields
increased by 50\% from SPS to RHIC naturally allow for a 20\% increase
of the transverse size and flow.  We also recall that the magnitude of
the elliptic flow coefficient, $v_2$, is similar from SPS to RHIC
\cite{voloshin}.

\item The similarity of the soft physics at RHIC and SPS may be
explained by the parton saturation phenomenon \cite{kgbw,iancu} (scenario
(2) of Ref. \cite{kln}). If the onset of saturation occurs already at
SPS energies, then the initial conditions for the multiparticle
production are similar at SPS and RHIC, explaining the similarities
discussed in this paper. The situation is reminiscent of the Hagedorn saturation
\cite{hagedorn,hag1,hag2} in elementary collisions, where the further increase
of the collision energy does not lead to increased temperature. 
\item The property of the saturation of soft physics, or at least the
very weak dependence on the collision energy, should help to verify
and constrain various models. The incomplete list of the most popular
approaches and ideas includes: thermal models
\cite{wbwf,hirsch,fbm,pbmsps,gaz1,raf,becatt,pbmrhic,gaz2},
hydrodynamic models
\cite{siemens,bjorken,baym,SSH,Rischke,SH,lcso,huo,tea,jap,heinzr},
transport theories \cite{dumitru}, and saturation models \cite{kln}.

\end{enumerate}


\begin{thebibliography}{99}

\bibitem{na44} I. G. Bearden et al., NA44 Collaboration, nucl-ex/0202019.

\bibitem{na49} P.~G.~Jones and the NA49 Collaboration,
Nucl. Phys. {\bf A610}, 188c (1996).

\bibitem{phenix} K. Adcox et al., PHENIX Collaboration, nucl-ex/0112006;
J. Velkovska, PHENIX Collaboration,  Nucl. Phys. {\bf A698}, 507 (2002).

\bibitem{stara} C. Adler et al., STAR Collaboration, Phys. Rev. Lett.
{\bf 87}, 262302 (2001).

\bibitem{starb} J. Harris, STAR Collaboration, contribution
to QM2001.

\bibitem{starc} C. Adler et al., STAR Collaboration, nucl-ex/0203016.

\bibitem{phenixhbt} K. Adcox et al., PHENIX Collaboration, nucl-ex/0201008.

\bibitem{starhbt} C. Adler et al., STAR Collaboration, Phys. Rev. Lett.
{\bf 87}, 082301 (2001).

\bibitem{agshbt} L. Ahle et al., E-802 Collaboration, nucl-ex/0204001. 

\bibitem{dumitru} S. Soff, S.~A.~Bass, and A. Dumitru, Phys. Rev. Lett.
{\bf 86}, 3981 (2001)

\bibitem{phobosmul} B.~B.~Back et al., PHOBOS Collaboration,
Phys. Rev. Lett. {\bf 85}, 3100 (2000).

\bibitem{phenixmul} K. Adcox et al., PHENIX Collaboration,
Phys. Rev. Lett. {\bf 87}, 052301 (2001).

\bibitem{na49mul} H. Appelsh\"auser et al., NA49 Collaboration,
Phys. Rev. Lett. {\bf 82}, 2471 (1999).

\bibitem{wa98} M.~M.~Aggarwal et al., WA98 Collaboration, Eur. Phys.
J. {\bf C18}, 651 (2001).

\bibitem{wbwf} W. Broniowski and W. Florkowski, Phys. Rev. Lett.
{\bf 87}, 272302 (2001).

\bibitem{hirsch} W.~Broniowski and W.~Florkowski, Proceedings of the
International Worksshop XXX on Gross Properties of Nuclei and Nuclear
Excitations, Hirschegg, Austria (GSI, Darmstadt, 2002), 146.

\bibitem{vitev} I. Vitev and M. Gyulassy, nucl-th/0104066.

\bibitem{michalec} M. Michalec, nucl-th/0112044.

\bibitem{fbm} W. Florkowski, W. Broniowski, and M. Michalec,
Acta Phys. Pol. {\bf B33}, 761 (2002).

\bibitem{bl} T. Cs\"{o}rg\H{o} and B. L\"{o}rstad, Phys. Rev. {\bf
C54}, 1390 (1996).

\bibitem{scheibl} R. Scheibl and U. Heinz, Phys. Rev. {\bf C59}, 1585
(1999).

\bibitem{strange} W. Broniowski and W. Florkowski, nucl-th/0112043.

\bibitem{PBMAGS}  P. Braun-Munzinger, J. Stachel, J. P. Wessels, and N. Xu,
Phys. Lett. {\bf B 344}, 43 (1995); Phys. Lett. B {\bf 365}, 1 (1996).

\bibitem{cest}  J. Cleymans, D. Elliott, H. Satz, and R. L. Thews, Z. Phys.
{\bf C74}, 319 (1997).


\bibitem{becattini}  F. Becattini, L. Bellucci and G. Passaleva,
Nucl. Phys. Proc. Suppl. {\bf 92}, 137 (2001).


\bibitem{voloshin} S.~A.~Voloshin, Proceedings of the International
Worksshop XXX on Gross Properties of Nuclei and Nuclear
Excitations, Hirschegg, Austria (GSI, Darmstadt, 2002), 207.

\bibitem{kgbw} K.~Golec-Biernat and M.~W\"usthoff, Phys. Rev. 
{\bf D59}, 014017 (1999); Phys. Rev. {\bf D60}, 114023 (1999); Eur. Phys.
J. {\bf C20}, 313 (2001).

\bibitem{iancu} E. Iancu, A. Leonidov, and L. McLerran,
Lectures given at the NATO Advanced Study Institute ``QCD
perspectives on hot and dense matter'', Carg\`ese, Corsica, France,
2001, hep-ph/0202270.

\bibitem{kln} D. Kharzeev, E. Levin, and M. Nardi, hep-ph/0111315.

\bibitem{hagedorn} R. Hagedorn, CERN preprint No. CERN-TH.7190/94 (1994),
and references therein.

\bibitem{hag1} W. Broniowski and W. Florkowski, Phys. Lett. {\bf B490},
223 (2000).

\bibitem{hag2} W. Broniowski, in Proc. of Few-Quark Problems, Bled, 
Slovenia, July 8-15, 2000, eds. B. Golli, M. Rosina, and S. \v Sirca, 
p. 14, hep-ph/0008122.

\bibitem{pbmsps} P. Braun-Munzinger, I. Heppe, and J.~Stachel,
Phys. Lett. {\bf B465 }, 15 (1999).

\bibitem{gaz1} M. Ga\'zdzicki and M. I.  Gorenstein, Phys. Rev. Lett.
{\bf 83}, 4009 (1999).


\bibitem{raf} J. Rafelski and J. Letessier,
Phys. Rev. Lett. {\bf 85}, 4695 (2000); hep-ph/0112027.

\bibitem{becatt}
F. Becattini, J. Cleymans, A. Keranen, E. Suhonen, and K. Redlich,
Phys. Rev. {\bf C64}, 024901 (2001). 

\bibitem{pbmrhic} P. Braun-Munzinger, D. Magestro, K. Redlich, and J. Stachel,
Phys. Lett. {\bf B518}, 41 (2001).

\bibitem{gaz2} K. A. Bugaev, M. Ga\'zdzicki, and M. I.  Gorenstein, 
Phys. Lett. {\bf B523}, 255 (2001).


\bibitem{siemens} P. J. Siemens and J. Rasmussen, Phys. Rev. Lett. {\bf 42},
880 (1979); P. J. Siemens and J. I. Kapusta, Phys. Rev. Lett. {\bf 43},
1486 (1979).

\bibitem{bjorken} J. D. Bjorken, Phys. Rev. {\bf D27}, 140 (1983).

\bibitem{baym} G. Baym, B. Friman, J.-P. Blaizot, M. Soyeur, and W.~Czy\.z,
Nucl. Phys. {\bf A407}, 541 (1983). 

\bibitem{SSH} E. Schnedermann, J. Sollfrank, and U. Heinz,
Phys. Rev. {\bf C48}, 2462 (1993).

\bibitem{Rischke} D. H. Rischke and M. Gyulassy, Nucl. Phys. {\bf
A697}, 701 (1996); Nucl. Phys. {\bf A608}, 479 (1996).

\bibitem{SH} R. Scheibl and U. Heinz, Phys. Rev. {\bf C59}, 1585
(1999).

\bibitem{lcso} A. Ster and T. Cs\"org\H{o}, hep-ph/0112064.

\bibitem{huo} P. Huovinen, P. F. Kolb, U. Heinz, P. V. Ruuskanen, and 
S. A. Voloshin, Phys. Lett. {\bf B503}, 58 (2001).

\bibitem{tea} D. Teaney, J. Lauret, and E. V. Shuryak,
Phys. Rev. Lett. {\bf 86}, 4783 (2001);  
Nucl. Phys. {\bf A698}, 479 (2002).

\bibitem{jap} T. Hirano, Phys. Rev. {\bf C65}, 011901 (2002);
T. Hirano, K. Morita, S. Muroya, and C. Nonaka, nucl-th/0110009.

\bibitem{heinzr} U. Heinz, Nucl. Phys. {\bf A661}, 140c (1999).

\end{thebibliography}
\end{document}